\documentclass[useAMS,usenatbib]{mn2e}
\usepackage{soul}
\setstcolor{red}
\usepackage{ulem}
\usepackage{float}
\usepackage{graphicx}
\usepackage{color}
\usepackage{wasysym}
\usepackage[utf8]{inputenc}
\usepackage{chngpage}
\usepackage{amssymb,latexsym}

\usepackage[]{hyperref}
\hypersetup{colorlinks=true, linkcolor=red, citecolor=blue}
\title{Long-term evolution of anomalous X-ray pulsars and soft gamma repeaters }
\author[O. Benli \& \"{U}. Ertan]{O. Benli\thanks{E-mail: onurbenli@sabanciuniv.edu} \& \"{U}. Ertan\\
Sabanc\i\ University, 34956, Orhanl\i\, Tuzla, \.Istanbul, Turkey}

\begin{document}

\date{2016 January 28}
\maketitle


\def\la{\raise.5ex\hbox{$<$}\kern-.8em\lower 1mm\hbox{$\sim$}}
\def\ga{\raise.5ex\hbox{$>$}\kern-.8em\lower 1mm\hbox{$\sim$}}
\def\be{\begin{equation}}
\def\ee{\end{equation}}
\def\ba{\begin{eqnarray}}
\def\ea{\end{eqnarray}}
\def\m{\mathrm}
\def\d{\partial}
\def\R{\right}
\def\L{\left}
\def\a{\alpha}
\def\acold{\alpha_\mathrm{cold}}
\def\ahot{\alpha_\mathrm{hot}}
\def\Mdotstar{\dot{M}_\ast}
\def\Omegastar{\Omega_\ast}
\def\OmegaK{\Omega_{\mathrm{K}}}
\def\Omegastardot{\dot{\Omega}_{\ast}}
\def\Mdotin{\dot{M}_{\mathrm{in}}}
\def\Mdot{\dot{M}}
\def\Edot{\dot{E}}
\def\Pdot{\dot{P}}
\def\Pddot{\ddot{P}}
\def\Msun{M_{\astrosun}}
\def\Lin{L_{\mathrm{in}}}
\def\Lcool{L_{\mathrm{cool}}}
\def\Lacc{L_{\mathrm{acc}}}
\def\Ldiss{L_{\mathrm{diss}}}
\def\Rin{R_{\mathrm{in}}}
\def\rin{r_{\mathrm{in}}}
\def\rlc{r_{\mathrm{LC}}}
\def\rout{r_{\mathrm{out}}}
\def\rco{r_{\mathrm{co}}}
\def\Rout{R_{\mathrm{out}}}
\def\Ldisc{L_{\mathrm{disc}}}
\def\Lx{L_{\mathrm{x}}}
\def\Md{M_{\mathrm{d}}}
\def\cs{c_{\mathrm{s}}}
\def\dEb{\delta E_{\mathrm{burst}}}
\def\dEx{\delta E_{\mathrm{x}}}
\def\Bstar{B_\ast}
\def\Bb{\beta_{\mathrm{b}}}
\def\Be{\beta_{\mathrm{e}}}
\def\Rc{R_{\mathrm{c}}}
\def\rA{r_{\mathrm{A}}}
\def\rp{r_{\mathrm{p}}}
\def\Tp{T_{\mathrm{p}}}
\def\dMin{\delta M_{\mathrm{in}}}
\def\dM*{\delta M_*}
\def\Rstar{R_\ast}
\def\Teff{T_{\mathrm{eff}}}
\def\Tirr{T_{\mathrm{irr}}}
\def\Firr{F_{\mathrm{irr}}}
\def\Tcrit{T_{\mathrm{crit}}}
\def\P0min{P_{0,{\mathrm{min}}}}
\def\Av{A_{\mathrm{V}}}
\def\ah{\alpha_{\mathrm{hot}}}
\def\ac{\alpha_{\mathrm{cold}}}
\def\tc{\tau_{\mathrm{c}}}
\def\p{\propto}
\def\m{\mathrm}
\def\fast{\omega_{\ast}}
\def\Alfven{Alfv$\acute{e}$n}
\def\418{SGR 0418+5729}
\def\142{AXP 0142+61}
\def\ql{\textquoteleft}
\def\qr{\textquoteright}
\def\rzero{r_{\mathrm{0}}} 
\def\Bzero{B_{\mathrm{0}}} 
\def\Po{P_{\mathrm{0}}}
\def\Smax{\Sigma_{\mathrm{max}} }
\def\Szero{\Sigma_{\mathrm{0}} } 
\def\dr{$\Delta$r }
\def\rbb1{R_{\mathrm{BB1}}}
\def\rbb2{R_{\mathrm{BB2}}}
\def\nh{N_{\mathrm{H}}}
\def\Nsd{N_{\mathrm{sd}}}
\def\Nsu{N_{\mathrm{su}}}
\def\Ntot{N_{\mathrm{tot}}}
\def\Gogus{G\"{o}\u{g}\"{u}\c{s}}
\def\Caliskan{\c{C}al{\i}\c{s}kan}
\def\0526{SGR 0526--66}
\def\1810{XTE J1810--197}
\def\1841{1E 1841--045}
\def\1048{1E 1048.1--5937}
\def\1622{PSR J1622--4950}
\def\1900{SGR 1900+14}
\def\2259{1E 2259+586}
\def\0142{4U 0142+61}
\def\1822{Swift J1822.3--1606}
\def\0501{SGR 0501+4516}
\def\0100{CXOU J010043.1--721134}
\label{firstpage}

\begin{abstract}

We have investigated the long-term evolution of individual anomalous X-ray pulsars (AXPs) and soft gamma repeaters (SGRs) with relatively well constrained X-ray luminosity and rotational properties. In the frame of the fallback disc model, we have obtained the ranges of disc mass and dipole field strength that can produce the observed source properties. We have compared our results with those obtained earlier for dim isolated neutron stars (XDINs). Our results show that (1) the X-ray luminosity, period and period derivative of the individual AXP/SGR sources can be produced self-consistently in the fallback disc model with very similar basic disc parameters to those used earlier in the same model to explain the long-term evolution of XDINs, (2) except two sources, AXP/SGRs are evolving in the accretion phase; these two exceptional sources, like XDINs, completed their accretion phase in the past and are now evolving in the final propeller phase and still slowing down with the disc torques, (3) the dipole field strength (at the poles) of XDINs are in the $10^{11}-10^{12}$ G range, while AXP/SGRs have relatively strong dipole fields between $1-6 \times 10^{12}$ G, and (4) the source properties can be obtained with large ranges of disc masses which do not allow a clear test of correlation between disc masses and the magnetic dipole fields for the whole AXP/SGRs and XDIN population.

\end{abstract}


\section{Introduction} \label{Intro}

Young isolated neutron stars manifest themselves as members of different populations, namely  anomalous X-ray pulsars (AXPs), soft gamma repeaters (SGRs), dim isolated neutron stars (XDINs), rotating radio transients (RRATs), central compact objects (CCOs) and the so-called \ql high-magnetic-field\qr~radio pulsars. The reason for these single neutron stars to emerge as different classes is likely to be the differences in their initial conditions, while some of these sources could be evolving in different phases of similar evolutionary tracks.

The initial condition of an isolated neutron star evolving in vacuum could be defined by its initial period, $P_0$, and the magnetic moment $\mu = \Bzero R^3 / 2$ where $R$ is the radius of the star, and  $\Bzero$ is the magnetic dipole field strength on the pole of the star. The rotational evolution of these sources are governed by the magnetic dipole torques. Using the dipole torque formula, $\Bzero$ could be estimated as $\Bzero \simeq 6.2 \times 10^{19} ~(P \Pdot)^{1/2}$ where $P$ and $\Pdot$ are the rotational period and the period derivative of the star. The surface emission is powered by the intrinsic cooling of the star \citep{Page_etal_06}. The cooling curves of the sources are found to be similar for crustal fields less than $\sim 10^{14}$ G, while for stronger fields the field decay becomes important, and significantly modify both the X-ray luminosity and the rotational evolution of the sources \citep{Vigano_etal_13}. In recent years, the magnetar model which was originally proposed to explain the properties of AXP/SGRs (see \citealt{Olausen_Kaspi_14} for a recent catalogue of AXP/SGRs\footnote{http://www.physics.mcgill.ca/~pulsar/magnetar/main.html}) was developed to explain the long-term rotational and X-ray luminosity evolution of the young neutron star populations in general. In this model, the dipole field strength on the star inferred from the dipole formula is  greater than $10^{14}$ G for most AXP/SGRs, and in the  $\sim 10^{13}$--$10^{14}$ G range for the XDINs. In the efforts to explain the properties of young neutron star systems in a single picture, the magnetar model has some difficulties in producing the X-ray luminosity, period and period derivatives of the individual sources simultaneously (\citealt{Vigano_etal_13}, see \citealt{Mereghetti_etal_15} for a recent review).        

The properties and likely evolutionary paths of a neutron star that evolves with a fallback disc are very different from those of the sources evolving in vacuum described above. In the fallback disc model (\citealt{Chatterjee_etal_00, Alpar_01}), the rotational history of a neutron star is determined mainly by the disc torques that are much stronger than the dipole torques in most cases. The X-ray luminosity is produced by mass accretion on to the star from the disc, or by intrinsic cooling of the star when accretion is not possible. It was suggested by \cite{Alpar_01} that the properties of different neutron star populations could be explained if the properties of fallback discs are included in the initial conditions in addition to $B$ and $P_0$. This model was developed to include the effects of inactivation of the disc at low temperatures and the X-ray irradiation of the disc, with the contribution of the cooling luminosity,  on the long-term evolution of the sources (\citealt{Ertan_etal_09, Caliskan_etal_13, Ertan_etal_14}). The evolutionary curves obtained from this model  are significantly different from those given by the analytical fallback disc solutions which cannot take dynamical outer disc radius and irradiation effects into account.    

This model that can explain the general X-ray luminosity, $\Lx$, and rotational properties of AXP/SGRs~\citep{Ertan_etal_09} with conventional dipole fields ($\sim 10^{12}$ G) was also extended to explain the long-term evolution of the so-called \ql high-B radio pulsar\qr~PSR J1734--333 with very low braking index ($n = 0.9 \pm 0.2$) \citep{Caliskan_etal_13} and recently discovered \ql low-B magnetars\qr~\418~and \1822~\citep{Alpar_etal_11, Benli_etal_13}. We should note here that the braking indices smaller than $n = 3$ could also be achieved by isolated neutron stars. This requires that the magnetic dipole field of the neutron star is buried into the inner crust by a post-supernova accretion of matter with total mass $10^{-3}-10^{-4} \Msun$. Subsequent re-emergence of the field leads the neutron star to a rotational evolution with a growing dipole field on a time-scale of $\sim 1 -100$ kyr. During this growing field phase, the braking index of the star could be significantly smaller than the conventional value, $n=3$, which is expected for the stars that spin down by the dipole torques of a constant magnetic dipole fields (see e.g. \citealt{Geppert_etal_99}, \citealt{Pons_etal_12}).

The number of sources with average and seemingly extreme properties are increasing with the new detections like the recently measured rapid and persistent change in the braking index of PSR J1846$-$0258. Limitations of the models could be understood through source-by-source analyses of each population. This systematic approach could also help to determine the differences in the initial conditions of different populations together with their likely evolutionary connections. Recently, we performed a detailed study for XDINs in the fallback disc model \citep{Ertan_etal_14}, and showed that the X-ray luminosity, period and period derivative of each XDIN source could be acquired simultaneously by neutron stars with fallback discs for certain ranges of the disc-mass and the dipole field strength. In the present work, we perform the same analysis for AXP/SGR sources with relatively well-known quiescent state properties, and compare our results with those obtained for XDINs by \cite{Ertan_etal_14}.    

The rotational powers of AXP/SGRs are much lower than the observed X-ray luminosities ($\sim 10^{33}$--$10^{36}$ erg s$^{-1}$) for most sources. They show short super-Eddington ($\leq$1 s) soft gamma-ray bursts. Among young neutron star groups, these sources have relatively high period derivatives ($\sim10^{-13}$ to $\sim10^{-10}$ s s$^{-1}$) and long periods, clustered in the 2--12 s range, like XDIN periods. AXP/SGRs are the most variable X-ray sources among the isolated neutron stars. Many sources were detected in the hard X-rays ($> 10$ keV) with luminosities comparable to the soft X-ray luminosities \citep{Kuiper_etal_04, den_Hartog_etal_06, Kuiper_etal_06}. In the fallback disc model, hard X-ray spectrum and the energy dependent pulse properties can be explained by the bulk motion and thermal Comptonization of the accreting matter in the accretion column (\citealt{Trumper_etal_10, Trumper_etal_13, Kylafis_etal_14}, \citealt{Zezas_etal_15}), with the properties consistent with the results of the long-term evolutionary models. These sources have also been observed in optical, near and mid infrared (IR) bands with spectra that can be well fitted by the emission spectrum of viscously active irradiated fallback discs (\citealt{Ertan_Caliskan_06, Ertan_etal_07}). Among these sources, AXP 4U 0142+61 was detected in a broad band from optical to mid--IR \citep{Wang_etal_06}. The model fits to this data provides an upper limit to $B_0$ which is less than $10^{13}$ G, also consistent with the fields estimated from the long-term evolution of these sources in the fallback disc model.

We briefly describe the model in Section~\ref{subsec:description of the model}. In Section~\ref{subsec:phases}, we illustrate the effect of the initial conditions (disc mass, magnetic dipole field and initial period) on the long-term evolution of the neutron stars. In Section~\ref{sec:source properties}, we summarize the observational properties of AXP/SGRs studied in the present work. Our results for twelve AXP/SGRs with a comparison to XDIN properties obtained by the same model are given in Section \ref{sec:results}. We discuss and summarize our conclusions in Section~\ref{sec:dis_cons}.

\section{Model} \label{sec:model}

\subsection{Description of the model} \label{subsec:description of the model}

We employ the fallback disc model developed by \cite{Ertan_etal_09}. We solve the disc diffusion equation with an initial surface density profile of a steady thin disc using the $\alpha$--prescription of the kinematic viscosity \citep{Shakura_Sunyaev_73}. At a given radial distance r from the star, the effective temperature of the disc is $\Teff \cong \sigma^{1/4} (D + \Firr)^{1/4}$ where $D$ is the viscous dissipation rate and $\Firr = (C \Mdot c^2)/(4 \pi r^2)$ is the irradiation flux, where $c$ is the speed of light, $C$ is the irradiation parameter which depends on the albedo and geometry of the disc. The dynamical outer radius, $\rout$, of the viscously active disc  corresponds to the current radial position of a critical minimum temperature, $\Tp$, for the disc to generate viscosity and be able to transport mass and angular momentum that is, $\rout = r(T=\Tp)$. In the long-term evolution, $\rout$ propagates inward with decreasing X-ray luminosity, $\Lx$, leaving inactive matter beyond $\rout$. Results of earlier work on the long-term evolution of AXP/SGRs and XDINs indicate that $\Tp \sim 100$ K, for a plausible range of irradiation efficiencies. In the simulations, we start with an initial $\rout$ $\sim 10^{14}$ at which the effective temperature of the disc is $\sim \Tp$ with the initial X-ray luminosity. The results obtained by \cite{Inutsuka_Sano_05} imply that the turbulent viscosity mechanism in the disc remains active for temperatures as low as $\sim 300$ K, which is in agreement with the critical temperatures we obtained from the model fits. 

In the fallback disc model, the main X-ray source of accreting AXP/SGRs is the accretion on to the poles of the star, which can be written as $\Lacc = G M \Mdot /R$ where $G$ is gravitational constant, $M$ and $R$ is the mass and radius of the star, $\Mdot$ is the mass flow rate on to the star. Intrinsic cooling luminosity, $\Lcool$, powers the star in the propeller phase or when the accretion luminosity decreases below $\Lcool$ in the late phases of evolution. We use the theoretical cooling luminosity curve calculated by \cite{Page_etal_06} for neutron stars with conventional dipole fields. When the accretion is not allowed, we calculate the total cooling luminosity considering also the energy dissipation due to magnetic and disc torques as suggested by \cite{Alpar_07}. In the fallback disc model, starting from the earlier works (e.g. \citealt{Chatterjee_etal_00, Alpar_01}), it is assumed that part of the inflowing disc matter is accreted onto the surface of the star in the spin-down phase. There are theoretical works (e.g. \citealt{Rappaport_etal_04} and \citealt{Dangelo_Spruit_12}), and strong observational evidences supporting this assumption. For instance, for the recently discovered transitional millisecond pulsars, it is estimated that the accretion – propeller  transition takes place at  accretion rates that are orders  of magnitude lower than the rate  corresponding to the spin up-down transition (\citealt{Archibald_etal_14}, \citealt{Papitto_etal_15}). The critical condition for the accretion--propeller transition is not well known. In our model, we assume that the mass-flow onto the star is allowed when the light cylinder radius, $\rlc$, is greater than the \Alfven~radius $\rA = (G M)^{-1/7}~\mu^{4/7}~\Mdotin^{-2/7}$, where $\mu$ is the magnetic dipole moment of the neutron star and $\Mdotin$ is the rate of mass-flow arriving the inner disc radius, $\rin$. 

The torque acting on the neutron star can be found by integrating the magnetic torque from $\rA$ to the corotation radius, $\rco = (G M / \Omegastar^2)^{1/3}$ (see \citealt{Ertan_Erkut_08} and \citealt{Ertan_etal_09} for details of the torque model). This assumes a wide boundary layer between $\rA$ and $\rco$, nevertheless most of the contribution to the spin-down torque comes from the radii close to $\rco$. Therefore, the magnetic torques integrated from $\rA$ to $\rco$, or across a narrow boundary layer, say from $2 \rco$ to $\rco$, give similar results. This is due to the very sharp radial dependence of the magnetic pressure ($\propto r^{-6}$). The total spin-down torque integrated from $\rA$ to $\rco$ can be written as, 
\be \label{eq1}
\Nsd = I ~\Omegastardot = \frac{1}{2} \Mdotin ~(G M \rA)^{1/2} ~(1 - \fast^2)
\ee  
where $I$ is the moment of inertia, $\Omegastardot$ is the rate of change of the angular velocity of the star, $\fast = \Omegastar / \OmegaK(\rA)$ is the fastness parameter, $\Omegastar$ is the rotational angular frequency of the neutron star and $\OmegaK(\rA)$ is the Keplerian angular velocity of the disc at the \Alfven~radius. Accretion from the inner disc on to the neutron star is allowed when $\rco<\rA<\rlc$. In this phase, we assume that $\Mdotin = \Mdot$, that is, we neglect losses through the propeller effect. This assumption does not affect our results qualitatively.

In the accretion phase, there is also a spin-up torque associated with the mass accretion on to the star from the co-rotation radius, $\sim \Mdot~(G M \rco)^{1/2}$. When $\fast^2 \gg 1$, this torque is negligible in comparison with the spin-down torque given in equation~\ref{eq1}. Nevertheless, some of the sources, especially those with relatively large disc masses, could approach the rotational equilibrium for a certain epoch of the evolution with decreasing efficiency of the net spin-down torque (Section~\ref{Intro}). In general, we can write down the total torque acting on the star as the sum of the spin-down and spin-up torques, $\Ntot = \Nsd + \Nsu$ where $\Nsu \cong \Mdot~(G M \rco)^{1/2}$. The magnetic dipole radiation also contributes to the spin-down torque, while the disc torque is usually the dominant spin-down mechanism. When $\Nsu$ is negligible, the total torque becomes equal to the spin-down torque given by equation~\ref{eq1}. In Section~\ref{sec:results}, we will show that most AXP/SGRs are in the accretion phase with $\fast^2 \gg 1$. We estimate that only two AXPs, namely 1E 2259+586 4U 0142+61, are evolving close to rotational equilibrium at present.
 
In our calculations, we neglect the magnetic field decay which could affect the evolution of AXP/SGRs in the accretion phase. From the magnetic field decay models (e.g. \citealt{Geppert_etal_99}, \citealt{Konar_Bhattacharya_99_2}), we estimate that this effect is not significant for most of these sources which have ages between $10^3$ and a few $10^4$ yr and dipole fields strengths $\sim 1-4 \times 10^{12}$ G (on the pole). Initial magnetic dipole fields of these sources could be a few times stronger than their field at present. The field could have decayed by a relatively large factor for the two AXPs, namely 1E 2259+586 and 4U 0142+61, which seem to evolve with relatively high disc masses. Our simplification puts an uncertainty on the ages of the model sources. The ages indicated by our model calculations could be taken as an upper limit to the actual ages of the sources. Nevertheless, incorporating a detailed field decay model in our calculations would not change the current field strengths of the sources estimated in the present work. After we complete our study on different neutron star systems with our current simple model,  a detailed population analysis of the young neutron star systems in the fallback disc model, considering also the effects of the field decay on the model curves, will be presented in an independent work. 

For AXP/SGRs and XDINs, our results are not sensitive to the initial period $P_0$ (see e.g. \citealt{Ertan_etal_09}). With different $P_0$ values, model curves converge to the same source properties in the long-term, provided that the sources enter the accretion phase in the early phases of their evolution. The minimum temperature of the viscously active disc, $\Tp$, is a basic disc property that is expected to be the same for all the fallback discs around young neutron star systems, even if the disc composition is not well-known. Following, \cite{Ertan_etal_09}, we keep $\Tp$ in the $\sim 50-150$ K range in the models of all different sources like low--B magnetars, \ql high-B radio pulsars\qr, and XDINs to remain self consistent. Similarly, we fixed the other main disc parameters, namely the irradiation strength $C = 1 \times 10^{-4}$ and the viscosity parameter $\alpha = 0.045$ following the results of \cite{Ertan_Caliskan_06} and \cite{Caliskan_Ertan_12}. The actual disc mass, $\Md$, and the magnetic field strength, $B_0$, at the pole of the star could differ from source to source. We repeat the simulations many times, tracing the values of $\Md$ and $B_0$, to find the ranges of these quantities that can reproduce the source properties. The disc mass, $\Md$, is estimated by integrating the initial surface density profile from the inner to the initial outer disc radius ($\sim 5 \times 10^{14}$ cm). 

\subsection{Long-term Evolutionary Phases of a Neutron Star with a Fallback disc} \label{subsec:phases}

In our model summarized in Section~\ref{subsec:description of the model}, there are four basic parameters that determine the evolutionary curves of the model sources: magnetic dipole field strength on the pole of the star, $\Bzero$, the initial disc mass, $\Md$, the initial period of the star, $P_0$, and the minimum critical temperature, $\Tp$, below which the disc becomes passive. The critical temperature $\Tp$ is degenerate with the irradiation parameter $C$, which is constrained to $\sim$1--7$ \times 10^{-4}$ range with the results obtained by \cite{Ertan_Caliskan_06}.

\begin{figure}
\centering
\includegraphics[width=.5\textwidth,angle=0]{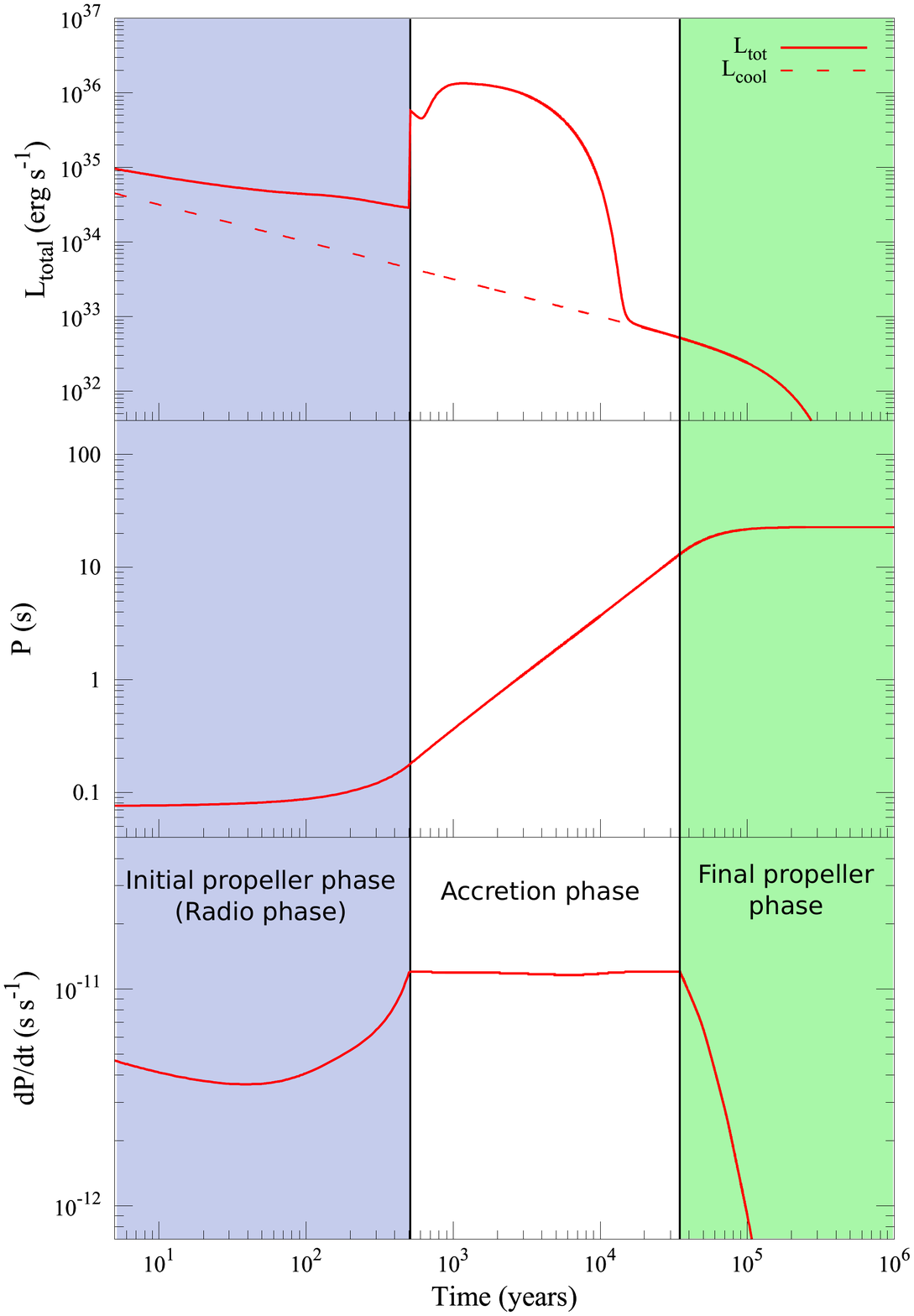}
\caption{A sample model curve that shows three basic evolutionary episodes of a neutron star-fallback disc system. For this illustrative model, $\Bzero = 2 \times 10^{12}$ G, $\Md = 5.7 \times 10^{-4}$ M$_{\astrosun}$ g cm$^{-2}$, $P_0 = 75$ ms and $\Tp = 150$ K. Durations of initial propeller phase (blue region), accretion phase (white region) and final propeller phases (green region) depend on the initial conditions. To reach the long periods of several seconds, a source should pass through the accretion phase. The initial propeller phase, which could be experienced by a fraction of the sources, has not a significant effect on the properties achieved in the accretion and final propeller phases (see the text for details).} 
\label{fig:phases}
\end{figure}

Depending on the initial parameters, the model sources could follow rather different evolutionary paths. The X-ray luminosity and rotational evolution of a source could pass through three basic evolutionary phases, namely the initial propeller phase, accretion phase and final propeller phase as seen in Fig.~\ref{fig:phases}. The top panel shows the X-ray luminosity, $\Lx$, evolution of the source. The abrupt rise in $\Lx$ at $t \simeq 5 \times 10^2$ yr is due to penetration of the inner disc into the light cylinder and the onset of accretion phase. This might happen at different times of evolution for different sources. Some sources may never enter the initial propeller phase, while some others remain always in this phase as radio pulsars depending on the initial conditions. The dashed line in the top panel represents the cooling history of a neutron star with a dipole field strength of ~$10^{12}$ G on the surface of the star. It is seen that $\Lcool$ dominates $\Lacc$ when accretion is not allowed or in the late phase of the accretion episode. The evolution is easier to follow from the period derivative, $\Pdot$, curve. The torque acting on the star is most efficient in the accretion phase. When the positive term is negligible, that is, when $\rA$ is not very close to $\rco$, $\Pdot$ is found to be independent of both $\Mdot$ and $P$. This constant $\Pdot$ behaviour of the illustrative model source in the accretion phase is seen in the bottom panel of Fig.~\ref{fig:phases}.

With decreasing $\Mdotin$, $\rA$ moves outward faster than the light cylinder radius, $\rlc$. In the model, accretion is switched off when $\rA$ is found to be greater than $\rlc$. For the illustrative model in Fig.~\ref{fig:phases} this correspond to $t \simeq 3 \times 10^4$ yr. From this point on, $\Pdot$ decreases with decreasing $\Mdotin$. In this final propeller phase, the sources do not accrete but still spin-down by the disc torques. It is seen in the middle panel of Fig.~\ref{fig:phases} that $P$ remains almost constant in this phase because of decreasing torque efficiency after termination of the accretion episode.

In the initial and final propeller phases, there is no accretion on to the star, and the pulsed radio emission is allowed. Nevertheless, we expect that sources could show pulsed radio emission only in the initial propeller phase, since in the final propeller stage, the sources with conventional dipole fields and long periods are usually not capable of producing pulsed radio emission; they are already below the pulsar death--line. Note that known and non-detected XDINs, and their progenitors would be expected to be radio pulsars as well if they had indeed strong dipole fields ($B = 10^{13}-10^{14}$ G) inferred from their observed $P$ and $\Pdot$. In the fallback disc model, XDIN properties could be achieved with $B \simeq 10^{12}$ G \citep{Ertan_etal_14} which together with long periods, place these sources under the pulsar death--line in the $B-P$ diagram.

\section{Source Properties} \label{sec:source properties}

\subsection*{SGR 0526--66:}
For SGR 0526--66, $P \simeq 8.1$ s and $\Pdot \simeq 3.8 \times 10^{-11}$ s s$^{-1}$ \citep{Tiengo_etal_09}. The 0.5--10 keV unabsorbed flux was reported as $\sim 10^{-12}$ erg s$^{-1}$ cm$^{-2}$ and for a distance of 50 kpc, which corresponds to LMC region, the bolometric X-ray luminosity of the source is $\sim 5 \times 10^{35}$ erg s$^{-1}$ \citep{Park_etal_12}.   

\subsection*{SGR 0501+4516:}
\noindent
For SGR 0501+4516, $P = 5.76$ \citep{Gogus_etal_08}, $\Pdot \simeq 5.8 \times 10^{-12}$ s s$^{-1}$ \citep{Gogus_etal_10}. Based on a likely association of the source with the supernova remnant HB9, \cite{Aptekar_etal_09} estimated the minimum distance to the source as 1.5 kpc. We have converted the observed X-ray flux into a bolometric X-ray luminosity range assuming that the source lies at a distance in the 1.5--5 kpc range. A black body + power-law model fit well to the quiescent soft X-ray spectrum of SGR 0501+4516 \citep{Camero_etal_14}. The best-fitting parameters are kT = $0.52 \pm 0.02$ keV with a blackbody radius of $0.39 \pm 0.0$5 km, $N_\mathrm{H} = 0.85(3) \times 10^{22}$ cm$^2$ and the power law index $\Gamma = 3.84 \pm 0.06$. This corresponds to an $\Lx$ range from $4.7 \times 10^{33}$ to $5.2 \times 10^{34}$ erg s$^{-1}$ for distances from 1.5 to 5 kpc.

\subsection*{XTE J1810--197:}
XTE J1810--197 has $P = 5.5$ s and $\Pdot \sim 4-10\times 10^{-12}$ s s$^{-1}$ \citep{Camilo_Cognard_etal_07}. The source showed transient pulsed radio emission during an X-ray outburst phase \citep{Halpern_etal_05}. It was the first pulsed radio detection from an AXP/SGR source. For a distance $d = 3.5$ kpc \citep{Bernardini_etal_09, Minter_etal_08}, converting the absorbed 0.6--10 keV data of \cite{Bernardini_etal_09} into 0.1-10 keV unabsorbed X-ray flux using the same spectral fit parameters, we estimate $\Lx \sim 1.6 \times 10^{34}$ erg s$^{-1}$. 

\subsection*{1E 1841--045:}
1E 1841--045 was discovered by \textit{Einstein HRI} \citep{Kriss_etal_85} and is the first source identified as an AXP with an X-ray pulse period of 11.8 s by \cite{Vasisht_Gotthelf_97}. This source is located in the supernova remnant Kes 73 with a distance of 8.5 kpc \citep{Tian_Leahy_08}. This source has the longest period in the currently known AXP/SGR population. A recent temporal analysis of the source gives $\Pdot \simeq 4.1 \times 10^{-11}$ s s$^{-1}$ \citep{Dib_Kaspi_14}. From the parameters of 0.5--10 keV spectrum in the quiescent state that is well described by black body + power-low with $\Gamma = 1.9 \pm 0.2$ and $kT = 0.43 \pm 0.003$ keV \citep{Kumar_Safi_10}, we estimate $\Lx \sim 5 \times 10^{35}$ with $d = 8.5$ kpc.

\subsection*{1E 1048.1--5937:}
A detailed temporal analysis of 1E 1048.1--5937 data, that were taken by \textit{RXTE} between 1996 July 3 and 2008 January 9 gives $P \simeq 6.5$ s and $\Pdot \sim 2.25 \times 10^{-11}$ s s$^{-1}$ \citep{Dib_etal_09}. The source showed fluctuations and anomalies during its temporal evolution for more than decades. We adopt $P$ and $\Pdot$ values measured by \cite{Dib_etal_09} in our analysis. Based on the X-ray spectral parameters obtained by \cite{Tam_etal_08}, we estimate $\Lx$ as $10^{35}$ erg s$^{-1}$ for a distance of $d \sim 9$ kpc \citep{Durant_vanKerk_06}.   

\subsection*{PSR J1622--4950:}
The PSR J1622--4950 was discovered in the radio band with a spin period of 4.3 s \citep{Levin_etal_10}. From the timing analysis of the radio pulses, \cite{Levin_etal_12} measured a factor of 2 decrease in the $\Pdot$ of the source since its discovery in radio. In the present work, we assume a $\Pdot$ range from 1 to 2 $\times 10^{-11}$ s s$^{-1}$. Single BB fits to the X-ray spectra of PSR J1622--4950, extracted from \textit{Chandra} and \textit{XMM-Newton} observations, at four different dates \citep{Anderson_etal_12}, give an unabsorbed flux $\sim 1.1 \times 10^{-13}$ erg s$^{-1}$ cm$^{-2}$ in 0.3--10 keV. From the dispersion measure, \cite{Levin_etal_10} gives a distance $\sim 9$ kpc. We estimate the bolometric X-ray luminosity $\sim 9.2 \times 10^{32}$ erg s$^{-1}$ for $d = 9$ kpc.       

\subsection*{SGR 1900+14:}
SGR 1900+14 has a period $P = 5.2$ s \citep{Hurley_etal_99, Kouveliotou_etal_99}. The most resent timing analysis indicated a spin-sown rate, $\Pdot = 9.2(4) \times 10^{-11}$ s s$^{-1}$ \citep{Mereghetti_etal_06}. The distance of the source is 12--15 kpc \citep{Vrba_etal_96}. We use the spectral fit parameters obtained by \cite{Mereghetti_etal_06}. For a hydrogen column density, $N_\mathrm{H} = 2.12 \times 10^{22}$ cm$^2$, converting the 2--10 keV unabsorbed flux into 0.1--10 keV unabsorbed flux by using WebPIMMS. we estimate the bolometric luminosity of the source $\Lx \sim 3 \times 10^{35}$ erg s$^{-1}$ for $d = 12.5$ kpc \citep{Davies_etal_09}.     

\subsection*{CXOU J010043.1--721134:}
The period and period derivative of CXOU J010043.1--721134 are $P = 8$ s and $\Pdot = 1.9 \times 10^{-11}$ s s$^{-1}$ \citep{McGarry_etal_05}. The source is located in Small Magellanic Cloud (SMC), dwarf satellite of our galaxy, with $d \sim 60$ kpc. The X-ray spectra of this source can be fit by a 2BB model with the parameters, $R_{\mathrm{BB1}} \sim 12.1$ km, $R_{\mathrm{BB2}} \sim 1.7$ km, $kT_1 \sim 0.3$ keV, $kT_2 \sim 0.7$ keV \citep{Tiengo_etal_08}. With $\nh = 5.9 \times 10^{20}$ cm$^2$ \citep{Dickey_Lockman_90} we obtain $\Lx \sim 6.5 \times 10^{34}$ erg s$^{-1}$.

\begin{figure}
\centering
\includegraphics[width=.7\textwidth,angle=270]{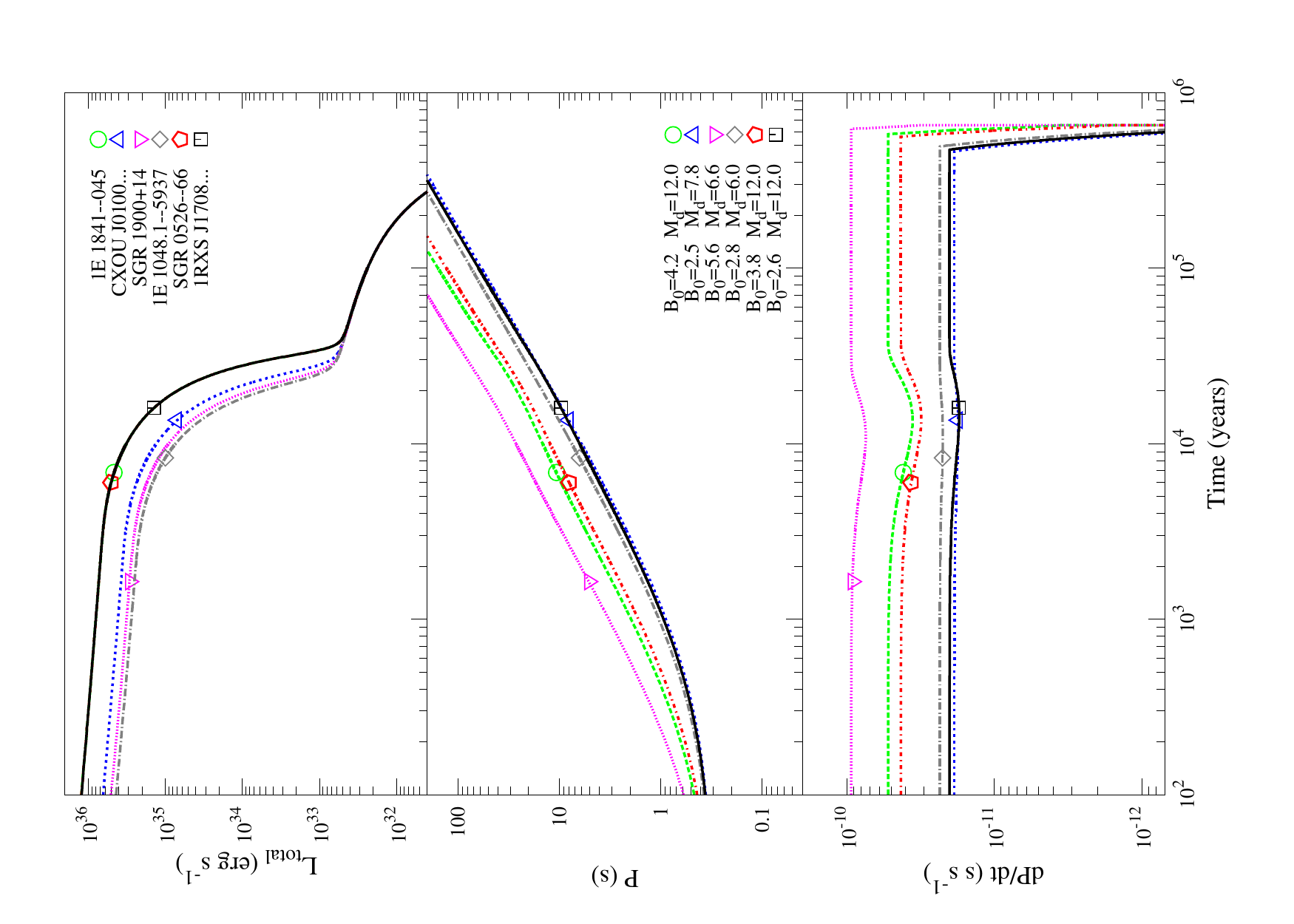}
\caption{The long-term evolutionary model curves of six AXP/SGRs. For all these AXP/SGRs, both $\Md$ and $\Bzero$ are constrained to very narrow ranges with central values given in the middle panel. The names of the sources are shown in the top panel. For these models, $\Tp = 100$ K and $C = 1 \times 10^{-4}$. $B_0$ and $\Md$ values are given in units of $10^{12}$ G and $10^{-6}$ M$_{\astrosun}$. For these sources accretion goes on till $t \sim 5 \times 10^5$ yr. But, the accretion luminosity remain below the cooling luminosity at $t \sim$ a few $10^4$ yr.
}
\label{fig:many in same} 
\end{figure}

\begin{figure}
\centering
\includegraphics[width=.7\textwidth,angle=270]{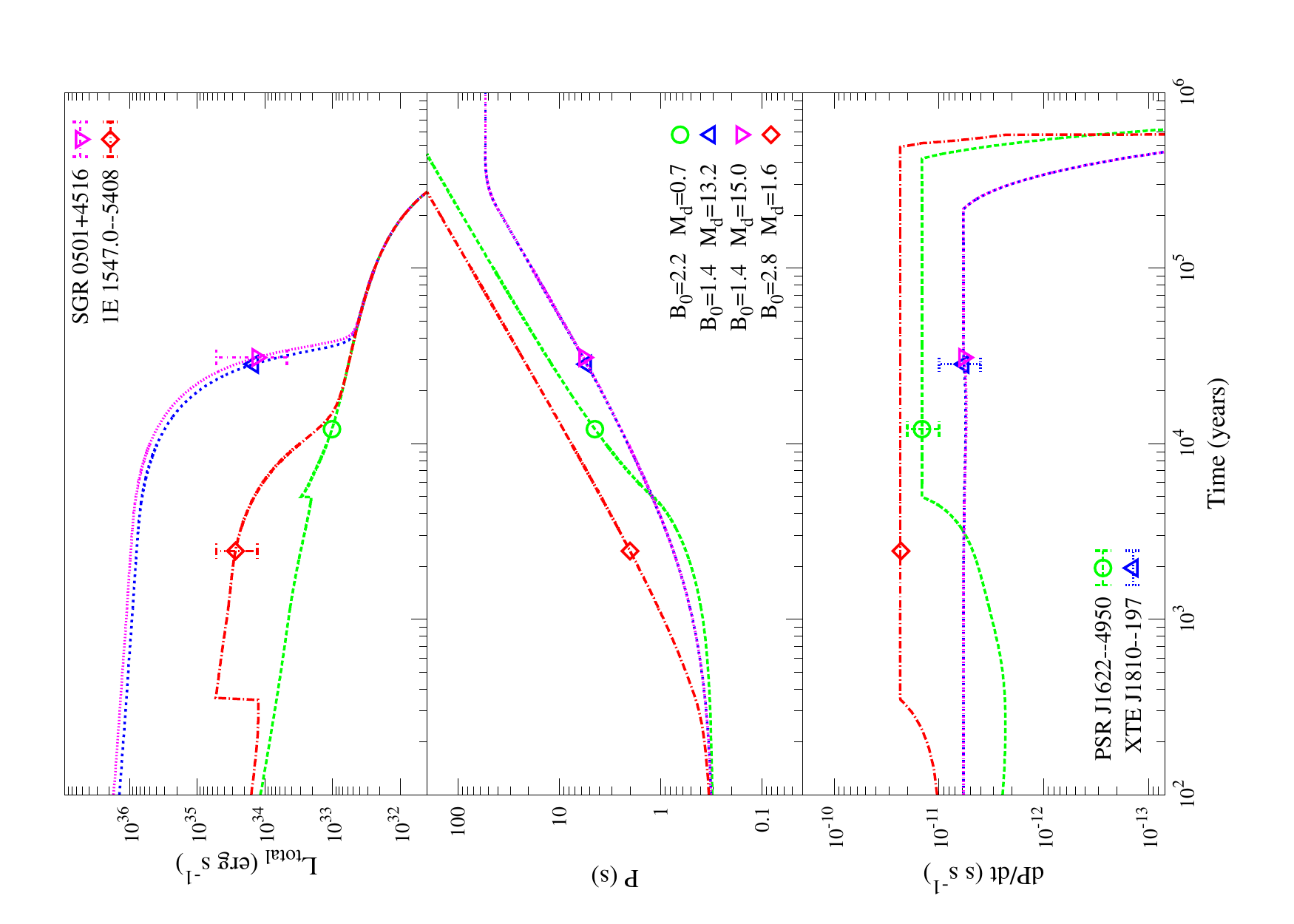}
\caption{The illustrative model curves representing the long-term evolutions of the four AXP/SGRs which have uncertainties in either $\Lx$ or $\Pdot$ measurements. The error bars show the uncertainties in measurements. For these sources, unlike the six sources given in Fig. \ref{fig:many in same}, our model cannot well constrain the $\Md$ and $\Bzero$ values. $B_0$ and $\Md$ values are given in units of $10^{12}$ G and $10^{-6}$ M$_{\astrosun}$. The model parameters are given in Table \ref{tab:AXP/SGRs}. The constant $\Pdot$ epochs correspond to accretion phases. 1E 1547.0--5408 and XTE J1810--197 enter the accretion phase at times $\sim 3 \times 10^2$ and $5 \times 10^3$ yr respectively. 
}  
\label{fig:errorbars} 
\end{figure}

\subsection*{1RXS J170849.0--400910:}
This source has $P = 11$ s, and $\Pdot = 1.9 \times 10^{-11}$ s s$^{-1}$ \citep{Dib_Kaspi_14}, and a distance of $3.8$ kpc \citep{Durant_vanKerk_06}. From the unabsorbed 0.5--10 keV flux reported by \cite{Rea_Israel_etal_07} we estimate $\Lx = 1.5 \times 10^{35}$ erg s$^{-1}$.
  
\subsection*{1E 1547.0--5408:}
This active transient source, also known as SGR J1550--5418, was discovered in 1980 by \textit{Einstein} satellite in the X-ray band. Later, the source was also observed in the radio band with a pulsation period, $P = 2.1$ s and $\Pdot = 2.3 \times 10^{-11}$ s s$^{-1}$ \citep{Camilo_Ransom_etal_07}. From the analysis of more resent X-ray observations with \textit{RXTE} and \textit{Swift}, its average period derivative was estimated as $\Pdot \simeq 4.8 \times 10^{-11}$ s s$^{-1}$ after the burst \citep{Dib_etal_12}. We adopt the $\Pdot$ value measured by \cite{Camilo_Ransom_etal_07} in the quiescent state. The unabsorbed flux in 0.5--10 keV band is $5.4 \times 10^{-12}$ erg s$^{-1}$ cm$^{-2}$ \citep{Bernardini_etal_11}. The X-ray luminosity is estimated to be $1.3 \times 10^{34}$ and $5.2 \times 10^{34}$ erg s$^{-1}$ for distances, $d = 4.5$ kpc (from the dust model by \citealt{Tiengo_etal_10}) and $d = 9$ kpc (derived from the dispersion measure by \citealt{Camilo_Ransom_etal_07}). We adopt this X-ray luminosity range for 1E 1547.0--5408 in our model calculations.

\subsection*{4U 0142+61:}
4U 0142+61 is one of the most extensively studied AXPs in different energy bands from optical to hard X-rays. It has $P = 8.7$ s and $\Pdot = 2 \times 10^{-12}$ s s$^{-1}$ \citep{Dib_Kaspi_14}. The soft($<10$ keV) and the hard X-ray luminosity of the source are $\sim 3.2 \times 10^{35}$ and $1.4 \times 10^{35}$ erg s$^{-1}$ respectively \citep{denHartog_etal_08}. In the quiescent state, 0.8--160 keV luminosity of the source is $\sim 4.6 \times 10^{35}$ erg s$^{-1}$ for a distance $d = 3.6$ kpc \citep{Durant_vanKerk_06}.  

\subsection*{1E 2259+586:}
This source has $P = 7$ s and $\Pdot = 4.8 \times 10^{-13}$ s s$^{-1}$ \citep{Dib_Kaspi_14, Morini_etal_88}. The unabsorbed flux of the source in 1--10 keV band is $5 \times 10^{-11}$ erg s$^{-1}$ cm$^{-2}$ in the quiescent state. For $d = 3.2$ kpc \citep{Kothes_etal_02}, the bolometric X-ray luminosity (including the hard x-ray emission) is assumed as $\sim 2 \times 10^{35}$ erg s$^{-1}$ in the quiescent state (see Table 3 of \citealt{Vigano_etal_13}).

\begin{figure}
\centering
\includegraphics[width=.7\textwidth,angle=270]{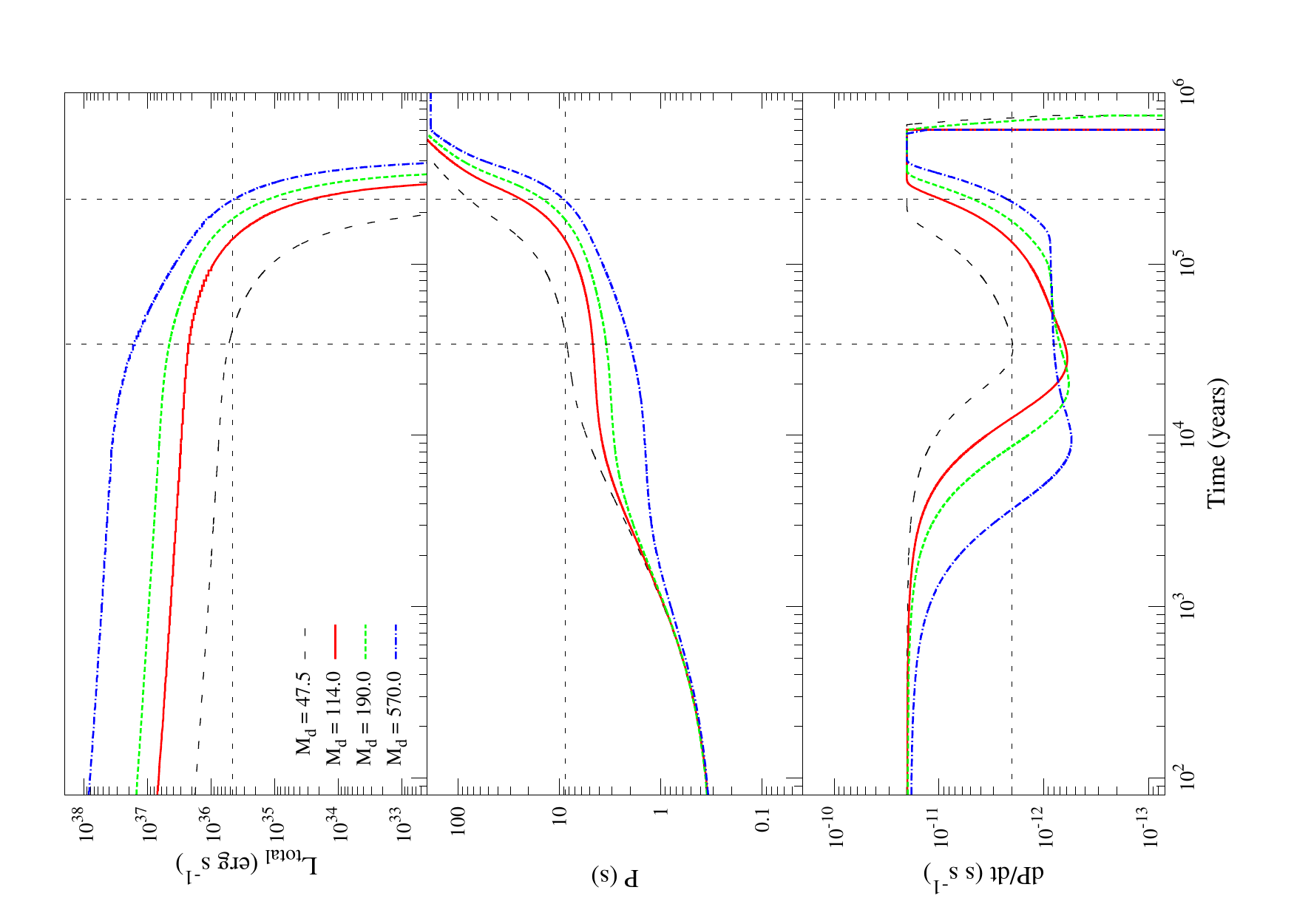}
\caption{Illustrative model curves that could represent the long-term evolution of 4U 0142+61. Reasonable model fits could be obtained with a large range of disc mass, a narrow $\Bzero$ range around $2.6 \times 10^{12}$ G, and $\Tp \sim 50-60$ K. These models are obtained with $\Tp = 54$ K and $\Md$ values (in units of $10^{-6}$ M$_{\astrosun}$) shown in the top panel. The model sources can acquire the source properties at the ages in the range limited by the vertical dashed lines shown in the figure ($\sim 3 \times 10^4$--$2 \times 10^5$ y). The horizontal dashed lines show the observed properties of the source. For these model curves, accretion remains as the dominant source of luminosity. }
\label{fig:0142} 
\end{figure}

\begin{figure}
\centering
\includegraphics[width=.7\textwidth,angle=270]{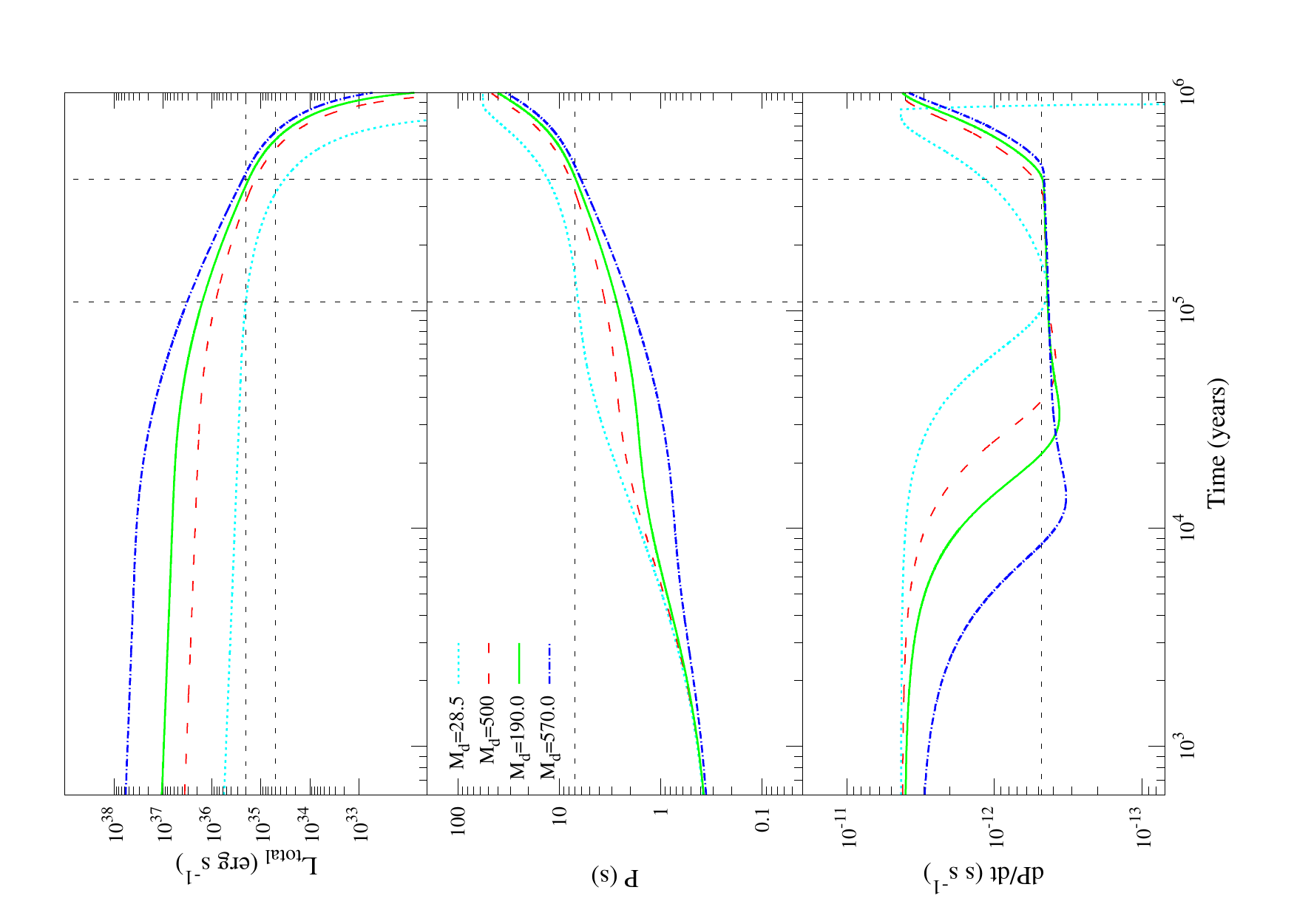}
\caption{The same as Fig. 7, but for 1E 2259+586. For all these model curves, $\Bzero = 1.2 \times 10^{12}$ G, $\Tp = 30$ K and $C = 1 \times 10^{-4}$. $\Md$ values are given in units of $10^{-6}$ M$_{\astrosun}$. Reasonable model fits for this source could be obtained with $\Tp \sim 30$--$40$ K at the ages $t \sim 1$--$4 \times 10^5$ yr (between the dashed vertical lines). These model sources, like those in Fig~\ref{fig:0142}, remain in the accretion phase their births to the present ages. } 
\label{fig:2259}
\end{figure}
\begin{table*} 
\footnotesize
\begin{minipage}{\linewidth}
\begin{center} 

\caption{The initial disc mass ($\Md$), dipole magnetic field strength at the pole of the star ($B_0$), minimum active disc temperature ($\Tp$) and the current age of the sources found from the models with the evolutionary curves given in Fig 4., and the observed properties $\Lx$, $P$ and $\Pdot$ of the sources. We take the irradiation parameter $C =10^{-4}$ and the viscosity parameter $\alpha = 0.045$ for all of the sources.}
\label{tab:AXP/SGRs}
\bigskip
\begin{tabular}{c|c|c|c|c|||c|c|c} \hline \hline 
Name     		& \begin{tabular}{@{}c@{}} $\Md$ \\ ( $10^{-6}$ M$_{\astrosun}$) \end{tabular}  & \begin{tabular}{@{}c@{}} $\Bzero$ \\ ($10^{12}$ G)\end{tabular} & \begin{tabular}{@{}c@{}} $\Tp$ \\ (K)\end{tabular} & \begin{tabular}{@{}c@{}} age \\ (yr)\end{tabular}	 & \begin{tabular}{@{}c@{}} $\Lx$ \\ ($10^{33}$ erg s$^{-1}$)\end{tabular} &\begin{tabular}{@{}c@{}} $P$\\  (s)\end{tabular}	& \begin{tabular}{@{}c@{}} $\Pdot$ \\ ($10^{-13}$ s s$^{-1}$) \end{tabular}  
\\ \hline 
SGR 0526--66    	&  	12.0		& 3.8		& 100 & $6 \times 10^{3}$				& 500			& 8		& 380		\\ 
XTE J1810--197  	& 6.0--24.0	& 1.2--1.8	& 100 &	2--4 $\times 10^{4}$ 			& 16			& 5.5	& 40--100	\\ 
1E 1841--045     	&   12.0		& 4.2		& 100 & $7 \times 10^{3}$				& $\sim$ 500	& 11.8	& 410		\\
1E 1048.1--5937 	&  	6.0		& 2.8		& 100 & $8 \times 10^{3}$				& 100			& 6.5	& 225		\\ 
PSR J1622--4950   	&  0.7--1.2	& 1.8--2.2	& 100 &	$\sim 1 \times 10^{4}$			& 0.9			& 4.3	& 100--200	\\ 
SGR 1900+14        	& 	6.6		& 5.6		& 100 & $1.6 \times 10^{3}$				& 300			& 5.2	& 920		\\ 	 
SGR 0501+4516		& 12.0--18.0	& 1.4		& 100 &	$3 \times 10^{4}$				& 4.7--52		& 5.76	& 58		\\
CXOU J0100...		& 	7.8		& 2.5		& 100 & $1.3 \times 10^{4}$				& 65			& 8		& 190		\\
1RXS J1708...		& 	12.0		& 2.6		& 100 & $1.4 \times 10^{4}$				& 150			& 11	& 190		\\
1E 1547.0-5408		& 1.0-2.1	    & 4.0		& 100 & $2 \times 10^{3}$				& 13--52		& 2		& 230		\\ \hline
\hline
4U 0142+61			& 47.5--570.0	& 2.4--2.6	& 54  & $\sim$ 0.3--2 $\times 10^{5}$	& 460			& 8.7	& 20		\\
1E 2259+586			& 28.5--570.0 & 1.2		& 30  & $\sim$ 1--4 $\times 10^{5}$	& $\sim$ 200	& 7		& 4.8		\\

\hline \hline 
\end{tabular}\\
\end{center} \end{minipage} \end{table*}


\begin{table*}
\caption{Observed properties of the six XDINs (see Ertan et al. 2014 and references therein for details). }
\label{tab:XDINs prop.}
\begin{minipage}{\linewidth}
\begin{center}
\bigskip
\begin{tabular}{c|c|c|c} \hline \hline
   &$P$ (s) & $\Pdot$ ($10^{-14}$ s s$^{-1}$) & $\Lx$ ($10^{31}$ erg s$^{-1}$)  \\ \hline
RX J0720.4--3125 & 8.39   & $\sim 7$	& $\sim 16$ \\
RX J1856.5--3754 & 7.06   & 2.97(7)	& 9.5		\\
RX J2143.0+0654  & 9.43   & 4.1(18)	& 11		\\
RX J1308.6+2127  & 10.31  & 11.2 		& 7.9		\\
RX J0806.4--4123 & 11.37  & 5.5(30)	& 2.5 		\\
RX J0420.0--5022 & 3.45	   & 2.8(3)		& 2.6 		\\
\hline \hline
\end{tabular}\\

\end{center}
\end{minipage}
\end{table*}

\begin{table*}
\caption{The disc parameters and the corresponding ages for the six XDINs. The viscosity parameter $\alpha = 0.045$ for all the model sources. (Taken from Ertan et al. 2014). }
\label{tab:XDINs param.}
\begin{minipage}{\linewidth}
\begin{center}
\bigskip
\begin{tabular}{c|c|c|c|c|c} \hline \hline
   & $B_{0}$ (10$^{12}$ G) & $\Md$ ($10^{-6}$ M$_{\astrosun}$) & $T_{p}$ (K) & $C$ (10$^{-4}$)& age (10$^{5}$ y)\\ \hline
RX J0720.4--3125 & 1.1--1.3   & 0.8--12.0   & 106 & 1   & 1.45 \\
RX J1856.5--3754 & 0.9--1.1   & 0.8--18.0   & 100 & 1   & 1.85 \\
RX J2143.0+0654  & 1.0--1.2   & 1.0--11.6   & 100 & 1   & 1.9 \\
RX J1308.6+2127  & 0.9--1.0   & 0.8--18.0   & 100 & 1.5 & 2.1 \\
RX J0806.4--4123 & 0.8--0.9   & 0.5--18.0      & 100 & 2.3 & 3.1 \\
RX J0420.0--5022 & 0.35--0.38 & 4.8--18.0     & 82  & 7   & 3.2 \\
\hline \hline
\end{tabular}\\

\end{center}
\end{minipage}
\end{table*}


\section{Results} \label{sec:results}

We could reproduce the observed source properties ($P$, $\Pdot$ and $\Lx$) of 12 AXP/SGRs in the frame of the fallback disc model (Figs. \ref{fig:many in same}, \ref{fig:errorbars}, \ref{fig:0142} and \ref{fig:2259}) using similar basic disc parameters ($\a = 0.045$, $C = 1 \times 10^{-4}$, $\Tp \sim 100$ K). We have chosen AXP/SGRs with relatively well-known quiescent-state properties for an investigation of the disc mass and magnetic field correlation. Our results indicate that the 12 AXP/SGRs that we have studied here are evolving in the accretion phase. Two sources, namely 4U 0142+61 and 1E 2259+586, seem to be the only AXP/SGRs that are currently evolving close to rotational equilibrium (see Figs~\ref{fig:0142} and~\ref{fig:2259}). We have obtained the model curves that can fit the individual source properties with the model parameters given in Table~\ref{tab:AXP/SGRs}. In the model, the dipole field strength, $\Bzero$, is well constrained for the accreting sources. Estimated $\Bzero$ and $\Md$ ranges given in Table~\ref{tab:AXP/SGRs} corresponds to the uncertainties in measured $\Pdot$ values (PSR J1622--4950, XTE J1810--197) and distances (SGR 0501+4516, 1E 1547.0-5408).  Our earlier results indicate that SGR 0418+5729 completed its evolution in the accretion phase, and is now evolving in the final propeller phase \citep{Alpar_etal_11}, Swift J1822.3--1606 could be in either accretion or in the final propeller phase depending on its actual X-ray luminosity in the quiescent state \citep{Benli_etal_13} and SGR 0501+4516 is found to be accreting from the fallback disc at present (\citealt{Benli_etal_15}).

Recently, \cite{Ertan_etal_14} analysed the individual source properties of six XDINs using the same model as we employ in the present work. With almost the same basic disc parameters, current X-ray luminosity and rotational properties of XDINs can also be produced by this model. For comparison with AXP/SGRs properties, we also present the results obtained by \cite{Ertan_etal_14} for XDINs in Table~\ref{tab:XDINs param.}. The properties of the six XDINs could be achieved by the sources in the final propeller phase, unlike most of the AXP/SGRs which seem to be still accreting matter from the disc (Figs. \ref{fig:many in same} and \ref{fig:errorbars}). In the propeller phases, the surface emission of the sources are powered mainly by the cooling luminosity. In the final propeller phase, we obtain the properties of XDINs with a large range of $\Md$. Nevertheless, $\Bzero$ values of these sources are well constrained despite the uncertainty in their initial disc masses. In Fig.~\ref{fig:correlations}, we plot the $B_0$ and $\Md$ values of both AXP/SGRs and XDINs, estimated from the model results. It is seen in Fig.~\ref{fig:correlations} that the magnitude of the dipole field on the pole of the star, $B_0$, ranges from $\sim 3 \times 10^{11}$ G to $\sim 6 \times 10^{12}$ G for the total AXP/SGR and XDIN sources. In addition, it is clearly seen that the $B_0$ values of XDINs ($\sim 3 \times 10^{11}$--$1 \times 10^{12}$ G) remains systematically below those of AXP/SGRs ($\sim 1 \times 10^{12} - 6 \times 10^{12}$ G). Due to large uncertainties in $\Md$, we could not determine whether there is a correlation between the disc mass and the dipole field strength of these sources.

\section{Discussion and Conclusions} \label{sec:dis_cons}

We have investigated the likely long-term evolutionary paths of individual AXP/SGR sources with relatively small uncertainties in  distances and period derivatives in the quiescent state. Our results show that the observed $P$, $\Pdot$ and $\Lx$ values of 14 sources, including the two low-B magnetars analyzed earlier \citep{Alpar_etal_11, Benli_etal_13}, could be achieved by neutron stars evolving with fallback discs and conventional dipole fields. Comparing Tables~\ref{tab:AXP/SGRs} and~\ref{tab:XDINs param.}, it is seen that the basic disc parameters, namely the irradiation strength, $C$,  the transition temperature between the active and passive disc state, $\Tp$, and the $\a$ parameter of the kinematic viscosity are all almost the same as the parameters used for XDINs in \cite{Ertan_etal_14}. The model can produce the individual source properties ($P$, $\Pdot$, $\Lx$) simultaneously for each of AXP/SGR and XDIN sources. 

Our results indicate that most AXP/SGRs are currently in the accretion phase. The two relatively old low-B magnetars seem to have completed the accretion epoch, and are currently spinning down by the disc torques without accretion on to the star, in the final propeller phase. Most of the accreting sources are not close to rotational equilibrium, with the exceptions 4U 0142+61 and 1E 2259+586 which seem to have initial disc masses greater than those of other AXP/SGR sources. We have obtained plausible model fits for these sources with relatively low $\Tp$ values (see Table~\ref{tab:AXP/SGRs}), which might be due to unknown details of the disc-field interaction geometry and efficiency when the \Alfven~radius approaches the co-rotation radius.

The birth rate of XDINs are estimated to be comparable to the radio pulsar birth rate and about an order of magnitude greater than that of AXP/SGRs (see e.g. \citealt{Popov_etal_06}). In other words, the number of currently known AXP/SGRs is not sufficient for these sources to be the \textit{only} progenitors of XDINs. Furthermore, our results imply that a fraction of AXP/SGRs have evolutionary paths not converging into the XDIN properties. In particular, some of AXP/SGRs could reach periods longer than the longest presently observed XDIN period, $\sim 12$ s, but at late phases with very low luminosities. One basic difference between these populations seems to be the relatively weak dipole fields ($\sim 10^{11}$--$10^{12}$ G) of XDINs (see Tables \ref{tab:AXP/SGRs} and \ref{tab:XDINs param.}). For this $\Bzero$ range, the observed upper limit for periods of XDINs is likely to be due to physical constraints rather than the selection effects (see Fig. 3 of \citealt{Ertan_etal_14}). Because, for given basic disc parameters, the dominant factor that determines the maximum period is the strength of the dipole field, rather than the disc mass. With the relatively weak dipole fields of XDINs, indicated by the model, the longest attainable periods are already close to the upper bound of observed periods. This result for the period upper limits needs more detailed investigation and should be taken with some caution due to our simplified condition for the transition between the accretion and propeller phases. The relation between $\Bzero$ and the period upper limit will be investigated in detail in an independent paper. In our model, the disc torque acting on the star is most efficient in the accretion epochs, and the increasing periods of the sources level off with termination of the accretion epoch. The results obtained by Ertan et al. (2014) imply that, unlike most AXP/SGRs, for the six XDIN sources\footnote{The period derivative of seventh XDIN has not been confirmed yet.} accretion phase terminated long ago, and they are slowing down in the final propeller phase at present. 

\begin{figure}
\centering
\includegraphics[width=.34\textwidth,angle=270]{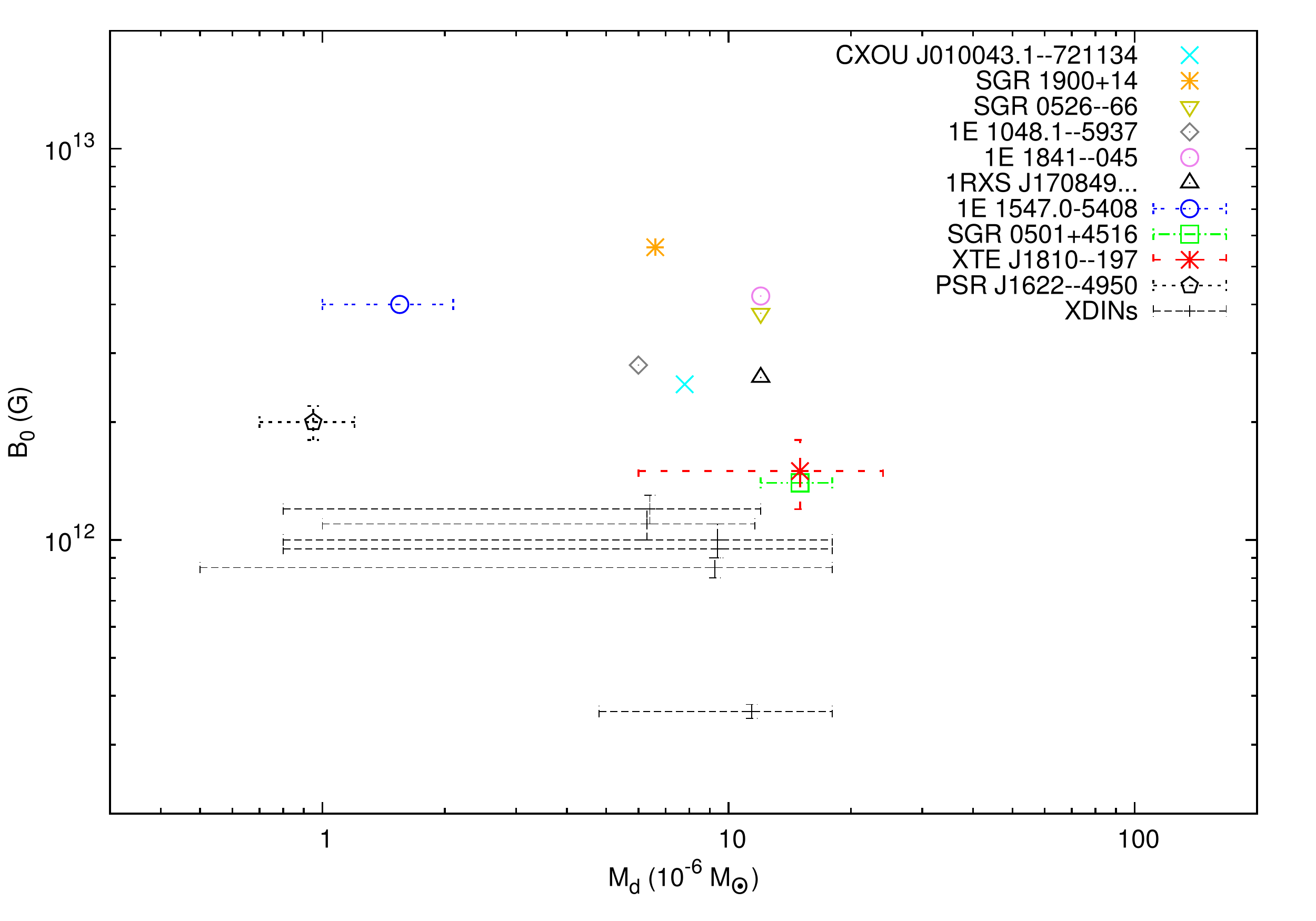}
\caption{The $\Bzero$ versus $\Md$ distribution of some AXP/SGRs (present work) and the six XDINs (from Ertan et al. 2014). The error bars represent the $\Bzero$ and $\Md$ ranges for which the model sources acquire the observed source properties $P$, $\Pdot$ and $\Lx$ simultaneously.} 
\label{fig:correlations}
\end{figure}

Since the model results do not well constrain the disc masses for XDINs, we are not able to determine whether there is a systematic relation between the actual disc masses and the dipole fields of the combined XDIN and AXP/SGR populations (see Fig.~\ref{fig:correlations}). Nevertheless, we estimate that XDINs are likely to be close to the lower bounds on their allowed disc masses seen in Fig. \ref{fig:correlations}, since evolution with low disc masses leaves most of the young XDINs below the current detection limits. This picture is consistent with the lack of detections of many young XDINs in X-rays at the ages of AXP/SGRs. This might also imply that the disc masses of most XDINs could form the peak of the initial disc mass distribution of the young neutron star systems that have fallback discs. Among the neutron star systems that evolve with fallback discs and can enter the accretion phase, those with stronger fields are more likely to have greater initial disc masses. Because, stronger fields require greater disc masses for the inner disc to enter the accretion phase. We could tentatively conclude that these systems that are located in the upper tails of both $B_0$ and $\Md$ distributions are likely to be the sources that are observed as X-ray luminous AXP/SGRs. In our model, $\Pdot \propto B_0^2$ in the accretion episode, and therefore among the accreting sources those with relatively high $\Pdot$ values have also stronger dipole fields. This conclusion cannot be generalized to the stars that are close to the rotational equilibrium, like 4U 0142+61 and 1E 2259+586 or the sources in the propeller phase, like XDINs.

As a final note, in our model, the sources in the first propeller phase could be observed as high-B radio pulsars. This is because the efficient disc torques increase $\Pdot$ significantly beyond the level that can be achieved by the magnetic dipole torques with the same field strength. A detailed population synthesis of these sources, which is beyond the scope of the present work, will be studied in a separate paper.

\section*{Acknowledgments}
We acknowledge research support from Sabanc\i\ University, and from T\"{U}B{\.I}TAK (The Scientific and Technological Research Council of Turkey) through grant 113F166. We thank M. A. Alpar for useful discussions and comments on the manuscript who is a member of the Science Academy--Bilim Akademisi, Istanbul, Turkey.

\bibliographystyle{mn2e}

\bibliography{mn-jour,benli}

\end{document}